\documentclass[sigconf]{acmart}

\AtBeginDocument{%
  \providecommand\BibTeX{{%
    \normalfont B\kern-0.5em{\scshape i\kern-0.25em b}\kern-0.8em\TeX}}}

\setcopyright{acmcopyright}
\copyrightyear{2018}
\acmYear{2018}
\acmDOI{XXXXXXX.XXXXXXX}

\acmConference[CIKM '24]{CIKM}{Oct 21--25,
  2024}{Boise}
\acmPrice{15.00}
\acmISBN{978-1-4503-XXXX-X/18/06}

\usepackage{booktabs}
\usepackage{enumitem}
\usepackage{xcolor}
\usepackage{subfigure}
\usepackage{multicol}
\usepackage{multirow}
\usepackage{ulem}

\usepackage{algpseudocode}
\usepackage{amsfonts}
\usepackage{amsmath}
\usepackage{amssymb}
\usepackage{algorithm}

\usepackage{makecell}
\usepackage{xspace}
\newcommand{\modelname}{ELCoRec\xspace}

\begin{document}

\title{Integrating Essential Supplementary Information into Large Language Models via Graph Attention for Recommendation}
\title{ELCoRec: Enhance Language Understanding with Co-Propagation of Numerical and Categorical Features for Recommendation}


\author{Jizheng Chen}
\authornote{Equal Contribution.}
\affiliation{%
  \institution{Shanghai Jiao Tong University}
  \city{Shanghai}
  \country{China}}
\email{humihuadechengzhi@sjtu.edu.cn}

\author{Kounianhua Du}
\authornotemark[1]
\affiliation{%
  \institution{Shanghai Jiao Tong University}
  \city{Shanghai}
  \country{China}}
\email{kounianhuadu@sjtu.edu.cn}

\author{Jianghao Lin}
\affiliation{
  \institution{Shanghai Jiao Tong University}
  \city{Shanghai}
  \country{China}}
\email{chiangel@sjtu.edu.cn}

\author{Bo Chen}
\affiliation{%
  \institution{Huawei Noah's Ark Lab}
  \city{Shanghai}
  \country{China}}
\email{chenbo116@huawei.com}

\author{Ruiming Tang}
\affiliation{%
  \institution{Huawei Noah's Ark Lab}
  \city{Shenzhen}
  \country{China}}
\email{tangruiming@huawei.com}

\author{Weinan Zhang}
\affiliation{%
  \institution{Shanghai Jiao Tong University}
  \city{Shanghai}
  \country{China}}
\email{wnzhang@sjtu.edu.cn}

\renewcommand{\shortauthors}{Kounianhua Du et al.}

\begin{abstract}
Large language models have been flourishing in the natural language processing (NLP) domain, and their potential for recommendation has been paid much attention to. Despite the intelligence shown by the recommendation-oriented finetuned models, LLMs struggle to fully understand the user behavior patterns due to their innate weakness in interpreting numerical features and the overhead for long context, where the temporal relations among user behaviors, subtle quantitative signals among different ratings, and various side features of items are not well explored. Existing works only fine-tune a sole LLM on given text data without introducing that important information to it, leaving these problems unsolved. 
In this paper, we propose \textbf{\modelname} to \textbf{E}nhance \textbf{L}anguage understanding with \textbf{Co}-Propagation of numerical and categorical features for \textbf{Rec}ommendation. Concretely, we propose to inject the preference understanding capability into LLM via a GAT expert model where the user preference is better encoded by parallelly propagating the temporal relations, and rating signals as well as various side information of historical items. The parallel propagation mechanism could stabilize heterogeneous features and offer an informative user preference encoding, which is then injected into the language models via soft prompting at the cost of a single token embedding. To further obtain the user's recent interests, we proposed a novel \textbf{R}ecent interaction \textbf{A}ugmented \textbf{P}rompt (\textbf{RAP}) template. Experiment results over three datasets against strong baselines validate the effectiveness of \modelname. The code is available at https://anonymous.4open.science/r/CIKM\_Code\_Repo-E6F5/README.md. 

\end{abstract}




\received{20 February 2007}
\received[revised]{12 March 2009}
\received[accepted]{5 June 2009}



\maketitle

\section{Introduction}
\label{sec:intro}
\begin{figure}[t!]
    \centering
    \includegraphics[width=\linewidth]
    {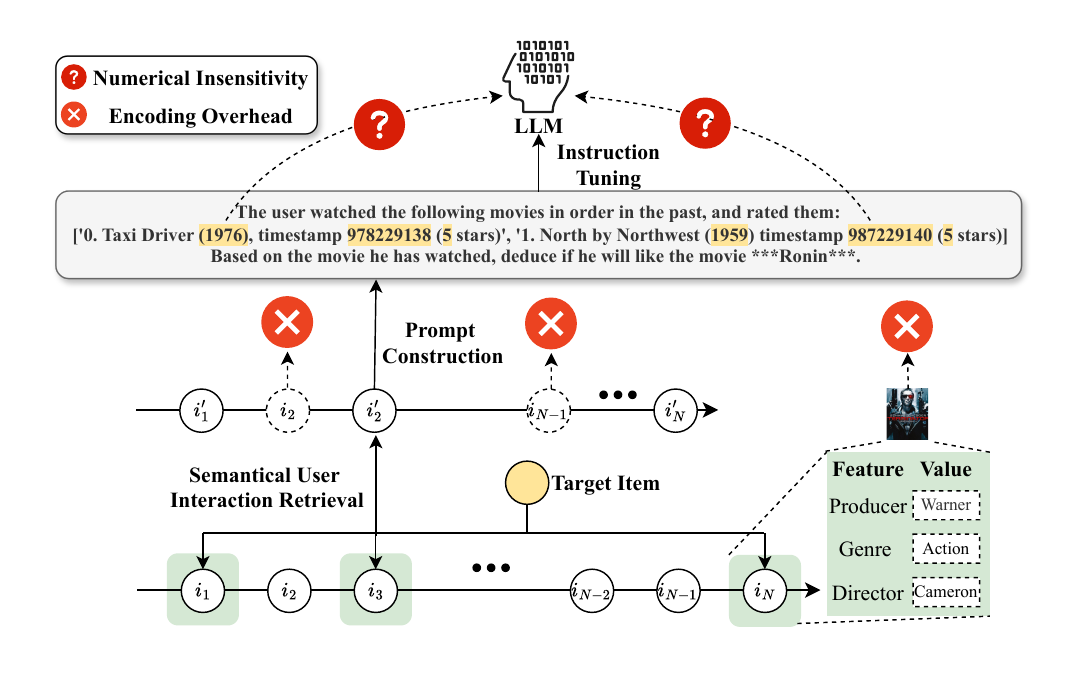}

    \vspace{-15pt}
    \caption{Illustration of the motivation. LLM struggles to understand numerical features such as rating and timestamps due to numerical insensitivity. Encoding overhead fails abundant side features such as producer and genre to form the template, and time series information is lost during retrieval.}
    \label{fig:intro}
\end{figure}
Recommender systems have permeated every corner of people's lives, including e-commerce~\citep{e-commerce}, streaming media~\citep{streaming1,din,dien}, and entertainment~\citep{entertainment} platforms. They efficiently filter the parts of vast amounts of data that users are interested in, thereby improving user engagement and company benefits. Besides, large language models (LLMs) have garnered widespread attention in the field of natural language processing (NLP) due to their outstanding reasoning and text-generation capabilities. Consequently, attempts to explore LLMs' potential for recommender systems emerge~\citep{LMRecSYS,zeroshotgpt,lin2023can}. \par
Nonetheless, LLMs have difficulties in comprehensively modeling user interests~\citep{usermodeldiff,rella}, hindering further performance improvements in recommendation. In this paper, we identify two main challenges in accurate user behavior modeling, as shown in Figure \ref{fig:intro}: (1) \textbf{Numerical insensitivity}. LLMs struggle to accurately understand the quantitative meaning of numerical information in prompt templates and treat them as normal tokens. For example, it may ignore the temporal relation among user behavior or overlook the precise quantification of user preferences provided by numerical ratings, instead treating them as plain textual strings. 
(2) \textbf{Encoding overhead}. LLMs have high inference latency and training costs, and their dialog windows have strict length limitations. This makes it unrealistic to construct prompt templates that encompass all abundant side features like genre and producer for the whole user sequence~\citep{rella}. 
Additionally, it necessitates filtering long user interaction histories to ensure the input response length is appropriate. Although user history retrieval has proved useful in extracting related items from whole user behavior sequence~\citep{rella,retrieval1,retrieval2}, valuable continuous time series information is lost due to the intermittent retrieved sub-sequence. These two issues result in the loss of a significant amount of information that is beneficial to recommendation performance during the process of constructing prompt templates from raw recommendation data.\par
Existing works haven't solved the above two problems efficiently. Early works~\citep{zeroshotgpt,bookgpt} only leveraged the reasoning capabilities of LLMs for zero-shot prediction, neglecting the critical information that LLMs struggle to understand in the input prompt. Some works~\citep{p5,ptab,tallrec,tabllm,rella} transform the recommendation task into a generative task via training LLMs on recommendation data.
However, this paradigm cannot deal with the numerical insensitivity problem. 
Recent works~\citep{e4srec,collm,clickprompt} introduce traditional click-through rate (CTR) prediction models 
such as SASRec~\citep{sasrec} and DCN~\citep{ dcnv2} to provide collaborative knowledge to LLMs. However, these works either: (1) directly apply existing general CTR prediction models without effectively capturing the relative numerical relation or (2) introduce excessive overhead of encoding side features for each historical item.\par

In this paper, we propose \textbf{\modelname} to \textbf{E}nhance \textbf{L}anguage understanding with \textbf{Co}-Propagation of numerical and categorical features to address the mentioned two problems for better \textbf{Rec}ommendation performance. To address the numerical insensitivity problem, numerical and categorical features are parallelly propagated by a GAT expert network to offer an informative user preference encoding that enhances LLM's understanding towards numerical features. To alleviate the the encoding overhead, the preference encoding is injected into the LLM's semantic space via soft prompting at the cost of a single token embedding. \modelname introduces a downstream graph attention network (GAT) as a unified expert network for feature encoding. GAT has unique advantages compared to other networks~\citep{wang2021propagate,hu2020heterogeneous,verma2019stability}, and we have meticulously designed the graph structure and feature construction approach. Compared to general CTR prediction models, our GAT expert network can better utilize heterogeneous nodes to encode features of different kinds. To capture the user's global interests while maximizing the recent information from the historical sequences, we further proposed a \textbf{R}ecent interaction \textbf{A}ugmented \textbf{P}rompt (\textbf{RAP}) template. Based on user history retrieval technology that filters out a sub-sequence of interacted items similar to the target item, RAP template additionally supplements the user's recent interaction history sequence to address the issue of continuous time series information loss caused by user history retrieval.\par

Our main contributions can be summarized as follows:

\begin{itemize}
    \item We identify two problems of LLMs, i.e., numerical insensitivity and encoding overhead, for user interest modeling. Further, we proposed to address the two problems simultaneously by co-propagation of numerical and categorical features via the unified GAT expert network. To the best of our knowledge, we are the first to identify and solve the two problems altogether in a unified framework.
    \item We proposed a novel framework \modelname, that injects informative user preference encoding generated by the unified GAT expert network via soft prompting at the cost of a single token embedding. A novel RAP template to better obtain user's recent interests is further proposed to form the textual input of \modelname.
    \item We conduct extensive experiments on three public datasets. Extensive experimental results demonstrate the outstanding performance and effectiveness of our model.
\end{itemize}

\section{Related Work}
\label{sec:relatedworks}
\subsection{Feature Interaction Methods}
Early CTR prediction methods focus on leveraging collaborative signals~\citep{map}. Classical FM~\citep{fm} learns a hidden vector for each feature, capturing second-order feature interactions. 
Based on it, FFM~\citep{ffm} introduces the concept of feature fields and divides each feature into different fields to capture combinatorial interaction relationships. Deep Crossing~\citep{deepcrossing} learns the interactions among features through deep neural networks. PNN~\citep{pnn} replaces the stacking layer with the proposed product layer to achieve more comprehensive feature interactions. Based on logistic regression, Wide\&Deep~\citep{widedeep} aims to enhance the model's memory and generalization capabilities with the wide part and the deep part, respectively. DCN~\citep{dcn} designed the cross network to replace the original wide part and introduced MoE structure along with low-rank operations in its improved version~\citep{dcnv2}. DeepFM~\citep{deepfm} replaces wide layer with an FM layer, enhancing the shallow model's feature interaction capabilities. xDeepFM~\citep{xdeepfm} uses a multi-layer CIN network to better capture high-order information.
\subsection{User Behavior Methods}
Recently more works leverage user behavior sequences to better model user interest~\citep{mamba4rec}. GRU4Rec~\citep{gru4rec} proposes to apply GRU~\citep{gru} units for sequential prediction. DIN~\citep{din} uses attention mechanisms to explicitly model the relevance between target items and user historical behaviors. DIEN~\citep{dien} models the evolution process of user interests using recurrent neural networks and captures the dynamic changes in user interests over time. SASRec~\citep{sasrec} applies self-attention mechanisms to user historical behavior sequences and applies position encoding to capture sequential relationships. DSIN~\citep{dsin} divides user behavior sequence into sessions to capture dynamic interests. Caser~\citep{caser} applies CNNs~\citep{cnn} to extract short-term preference sequence. MIMN~\citep{mimn} proposes to capture multiple aspects of user interests via a memory network architecture. SIM~\citep{retrieval1} leverages retrieval to model long user behaviors.  
\subsection{Semantic-Enhanced Methods}
Large language models (LLMs) are being researched to assist recommendation tasks~\citep{LMRecSYS,ptab,unitrec,PromptNR,RecFormer,tabllm,zeroshotgpt,Flant5,bookgpt,tallrec,glrec,CTR-BERT,kar,taste,clickprompt,collm,flip,e4srec,ubert}. CTRL~\citep{CTRL} aligns tabular data space with response space using contrastive learning. UniSRec~\citep{UnisRec} achieves unified cross-domain recommendation by learning item representations through a fixed BERT model. Building on it, VQ-Rec~\citep{VQ-Rec} applies vector quantization to learn better item representations. U-BERT~\citep{ubert} enhances user representation quality by encoding review information through the BERT model. LMRecSys~\citep{LMRecSYS} directly uses a pre-trained language model as a recommendation system. P5~\citep{p5} trains a large language model on the T5~\citep{t5} model for various downstream recommendation tasks. PTab~\citep{ptab} models tabular data via modality conversion, language model fine-tuning, and prediction model fine-tuning steps. TALLRec~\citep{tallrec} trains a large language recommendation model (LRLM) and designs a complete training paradigm and prompt templates for it. CoLLM~\citep{collm} proposed to provide collaborative knowledge to the large model using ID and interaction information. ReLLa~\citep{rella} trains the large language model on retrieved data, addressing the issue of long-sequence interest modeling.

\section{Preliminaries}
\subsection{Problem Formulation}
We focus on the click-through rate (CTR) prediction task, which is the core task in recommender systems to estimate a user’s click probability towards a target item given a certain context. 
For the $i$'s sample in the dataset, traditional CTR prediction task can be formulated as:
\begin{equation}
    \label{equation:ctr}
    p(y_i|\mathbf{X}^U_i, \mathbf{X}^I_i, \mathbf{X}^C_i, [\langle \mathbf{X}^I_{i_k}, y_{i_k}\rangle]_{k=1}^{K_i}, \theta),
\end{equation}
where $y_i$ denotes the prediction results ($0$ or $1$). $\mathbf{X}^U_i$ and $\mathbf{X}^I_i$ is the set of user features and item features, respectively. $\mathbf{X}^C_i$ is the corresponding context features (e.g. timestamp or weather), and $\theta$ denotes model parameters. $[\langle \mathbf{X}^I_{i_k}, y_{i_k}\rangle]_{k=1}^{K_i}$ represents the user behavior sequence, which contains $K_i$ items in total and we use $y_{i_k}$ to represent each history ratings.\par
For LLMs, prompt template construction is required to convert data sample $x_i$ into regular text $x_i^{text}$ followed by a binary question about the user preference towards the target item. The binary label $y_i$ also needs to be transformed into corresponding answers $y_i^{test} \in \{\mathbf{`` Yes"}, \mathbf{``No"}\}$. As we mentioned before, only the key features such as item title and a sub-sequence of user behavior items will be adopted to form a prompt template due to window length restrict and time concern. Auxiliary information can also be add to LLM, and finally the task is formulated as:
\begin{equation}
    \label{equation:ctr}
    p(y_i^{text}|x_i^{text}, I_x,\theta),
\end{equation}
where $I_x$ represents the external auxiliary information, $x_i^{test}$ is the textual prompt template.

\subsection{Graph Attention Networks}
Graph attetion networks~\citep{gat} (GATs) are commonly used to process graph-structured data. It combines the principles of Graph Neural Networks~\citep{gnn} (GNNs) with the attention mechanisms to more accurately and robustly learn user and item feature via dynamically assigning different levels of importance to different neighbors.\par
GAT works by first representing inputs as a topological graph structure. Each node $i$ in the graph has an initial feature vector $\mathbf{h}_i \in \mathbb{R}^d$. In the $l$'s graph attention layer, a feature aggregation process is first conducted by calculating the attention coefficient $e_{ij}$ for a given node $i$ and each of its neighbor $j$. $e_{ij}$ reflects the importance of node $j$’s features to node $i$, which is formulated as:
\begin{equation}\label{equ:gat attention eij}
    e_{ij} = a([\mathbf{W}\mathbf{h}_i^l\|\mathbf{W}\mathbf{h}_j^l]), j\in \mathcal{N}_i , 
\end{equation}
where $a$ is the activation function, and $\mathbf{W}$ is learnable parameter matrix. The normalized attention coefficients across all neighbors of a node can be calculated using a softmax function, denoted as:
\begin{equation}\label{equ:gat attention score}
\alpha_{ij} = \frac{\exp(LeakyReLU(e_{ij}))}{\sum_{k\in \mathcal{N}_i}\exp(LeakyReLU(e_{ik}))},
\end{equation}
where $LeakyReLU$ stands for the LeakyReLU activation function~\citep{leakyrelu}.\par
Afterwards, the feature vector of node $i$ is updated by aggregating the features of its neighbors weighted by the attention coefficients, which is formulated as:
\begin{equation}\label{equ:gat update}
\mathbf{h}_i^{(l+1)} = Linear(\mathop{\Vert}\limits_{k=1}^K\sigma(\sum_{j\in \mathcal{N}_i}\alpha_{ij}^k\mathbf{W}^k\mathbf{h}_j^{(l)})),
\end{equation}
where $\|$ denotes concatenation, $Linear$ is a linear layer mapping the vector to the embedding size. A multi-head attention mechanism is applied and $\alpha_{ij}^k$ and $\mathbf{W}^k$ stands for the attention coefficients and weight matrix of the $k$'s attention head, respectively. By feature aggregation and updating, GAT can effectively model the bidirectional interaction across graph nodes, thus can be applied to capture collaborative signal between user and interacted items. \par
In this work we view each user and item, as well as each categorical item feature as a node, and initialize the node embedding using feature encoding. Detailed explanation of our GAT expert network is in Section \ref{subsection: gat training}, and we have the whole propagation process of graph attention layer defined as:
\begin{equation}\label{equ:gat defination}
\mathbf{h}_i^{(l+1)} = f_{GAT}(\mathbf{h}_i^{(l)}, \{\mathbf{h}_j^{(l)}\}), j\in \mathcal{N}_i.
\end{equation}


\section{Methodology}
\begin{figure*}[!t]
    \centering
    \includegraphics[width=1.0\textwidth]{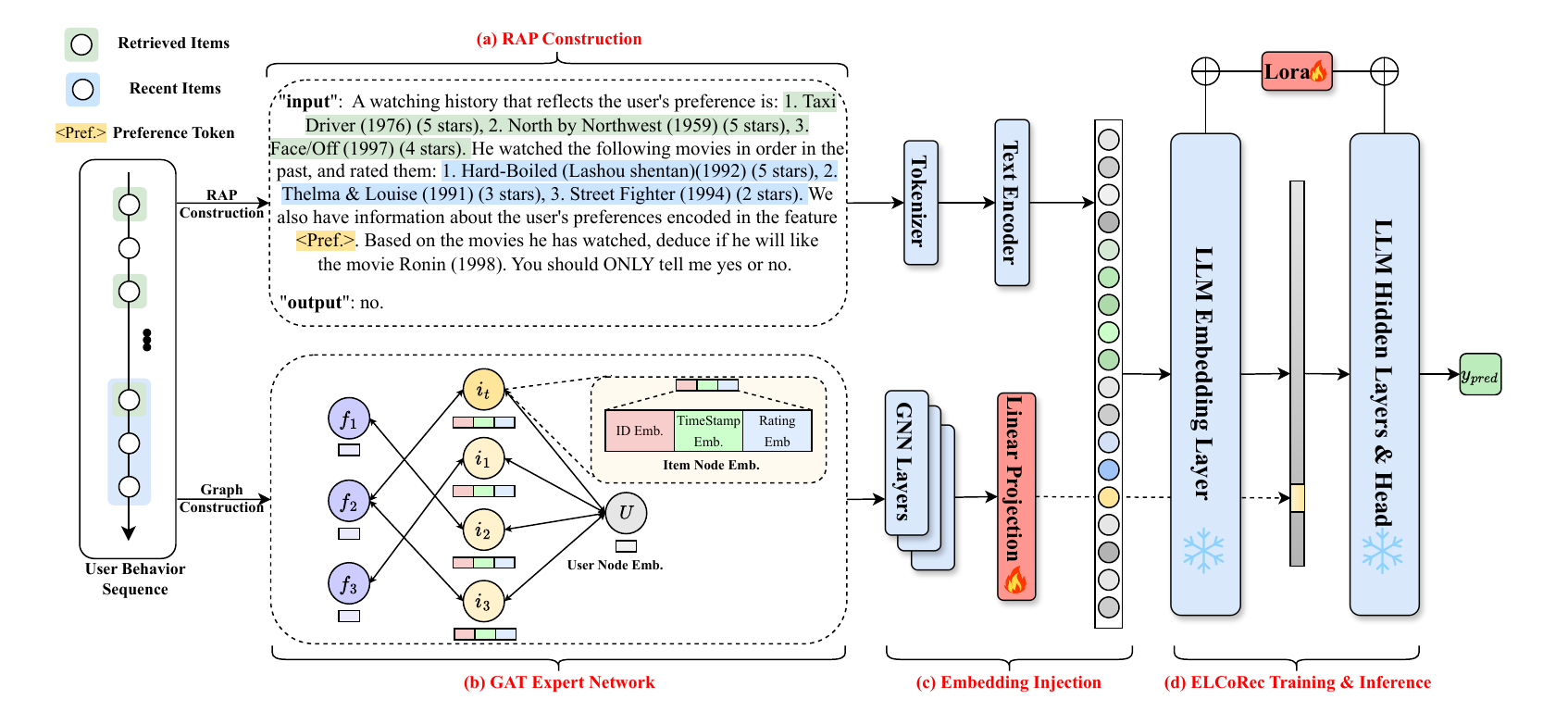}
    \vspace{-25pt}
    \caption{Framework overview. (a) RAP that connects user history retrieved item sequence (marked in green) and recent item sequence (marked in blue) along with the placeholder token for embedding injection (marked in orange), to form textual prompt. (b) GAT expert network. User and item' side features along with numerical infomantion are encoded and interacted. (c) Representations from GAT expert network are injected into LLM's latent space and (d) perform instruction tuning.}
    \label{fig:framework}
\end{figure*}

In this section, we describe the methodology of the proposed RAP template and \modelname framework in detail, which integrates extra beneficial information into large language models with downstream GAT expert network.

\subsection{Overview}
Our model framework is illustrated in Figure \ref{fig:framework}. To conduct user history retrieval based on semantic similarity in the construction of RAP prompt templates, we first encode the feature descriptions of all items using a large language model encoder. Specifically, the feature information of each item is formed into textual descriptions as field-value pairs, followed by being fed into a large language model to obtain semantic representations. To reuse the encoded information, we build a knowledge index library and pre-store the encodings of all items, with the item index as the key and corresponding semantic embedding as the value.\par
The workflow of our proposed model contains mainly four stages:\par 

\begin{itemize}[leftmargin=10pt]
    \item \textbf{RAP template construction.} For a given user interaction history sequence as input data, we first obtain two different item sub-sequences through retrieval enhancement and recent item retention, respectively. RAP template combines the obtained two sequences along with placeholders for subsequent semantic injections. This will form the textual input for the large language model.
    \item \textbf{GAT expert network training.} The GAT expert network is first pre-trained, and the user's interaction history will be encoded by a downstream GAT expert network to generate user's preference embedding towards the target item.
    \item \textbf{Embedding injection.} User preference representation is injected into the large language model's semantic space at the cost of one single token after being mapped to the same dimensionality through a linear layer.
    \item \textbf{LLM finetuning.} Instruction tuning is performed using LoRA~\citep{lora} on the injected representation.
\end{itemize}

In this manner, continuous time-series information is maximally preserved in the input template, and the large language model can also learn essential information which it can not inherently understand from the high-quality representations encoded by the GAT expert network.

\subsection{RAP Template}\label{subsection: rap construction}
\begin{figure}[h]
    \centering
    \includegraphics[width=\linewidth]{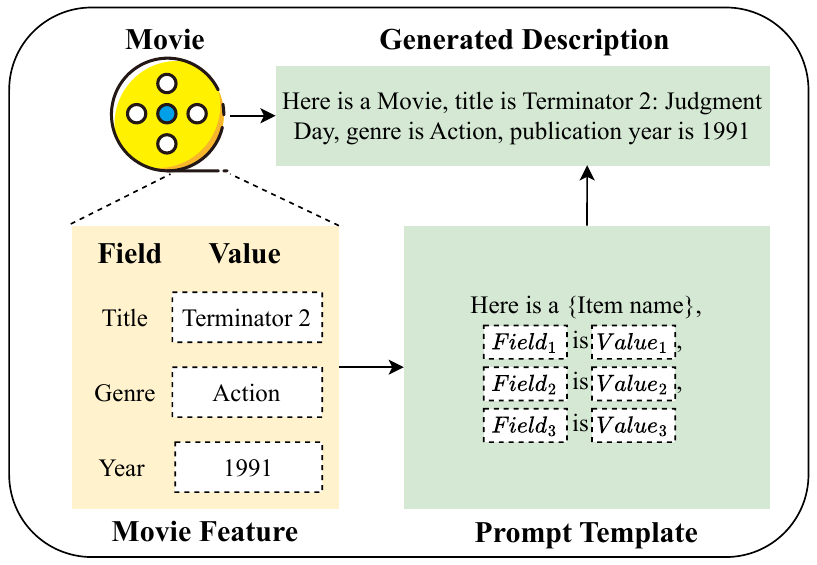}
    \vspace{-15pt}
    \caption{Item description construction. Each item's semantic description is obtained via the ``Feature is Value'' template. }
    \vspace{-15pt}
    \label{fig:item desc}
\end{figure}

Due to the context window limitation and sequence incomprehension problem~\citep{rella}, it is crucial to properly design the prompt template for LLMs to include the user's historical items that could largely reflect interest trends. To this end, we propose the \textbf{R}ecent Interaction \textbf{A}ugmented \textbf{P}rompt (RAP) template to capture both the global information related to the target item and the recent information emphasizing the latest trends of user preferences.\par

To capture the global information related to the target item, we propose to retrieve the semantically relevant behaviors from the user's historical sequence via vector matching. We first build a knowledge index library to efficiently extract and store the semantic representations of items in the dataset. Specifically, for each item, we construct a semantic description template using a key-value pair structure. As illustrated in the example of the MovieLens-1M movie recommendation scenario, the movie in the original dataset is named "Terminator 2: Judgment Day," with the genre as Action and the release year as 1991. According to the key-value pair template construction rules, generated item descriptions can be obtained, as shown in Figure \ref{fig:item desc}.

Next, we input the obtained textual descriptions of the items into large language model $\Phi_S(\cdot)$ to obtain their textual embedding representations. These representations are stored in an indexed knowledge base $KB[\cdot]$, which is indexed by item numbers as lookup keys and the textual embeddings of the items as values. The construction of this knowledge index not only allows for the efficient reuse of encoded text vectors but also makes the process of similarity calculation orderly and standardized.\par
Previous works~\citep{retrieval1,retrieval2} indicate that user history retrieval can purify and denoise historical sequences, allowing the large language model to better capture the user's interest concerning on specific target items. Concretely, given target item $i_t$ along with user history item sequence $i_1,i_2,...,i_N$, we first acquire their semantic embedding from constructed knowledge base, denoted as $i_t^s$ and $i_1^s, i_2^s,...,i_N^s \in \mathbb{R}^d$, respectively. $d$ represents LLM's embedding dim, and since we use Vicuna-13B~\citep{vicuna}, a text generation model based on Transformer\cite{attention} structure, we take the vectors from the last hidden layer as the semantic embedding representations. Subsequently, we measure the semantic relevance of each historical item to the target item using cosine similarity, and retrieve the most similar $K$ items, denoted by:
\begin{equation}\label{equ: retrieval results}
    \{x_i^{text}\}_{i=1}^K = \max \{ cos\_sim(i_t^s, i_j^s)\}_{j=1}^K.
\end{equation} 

\begin{figure}[h]
    \centering
    \includegraphics[width=\linewidth]{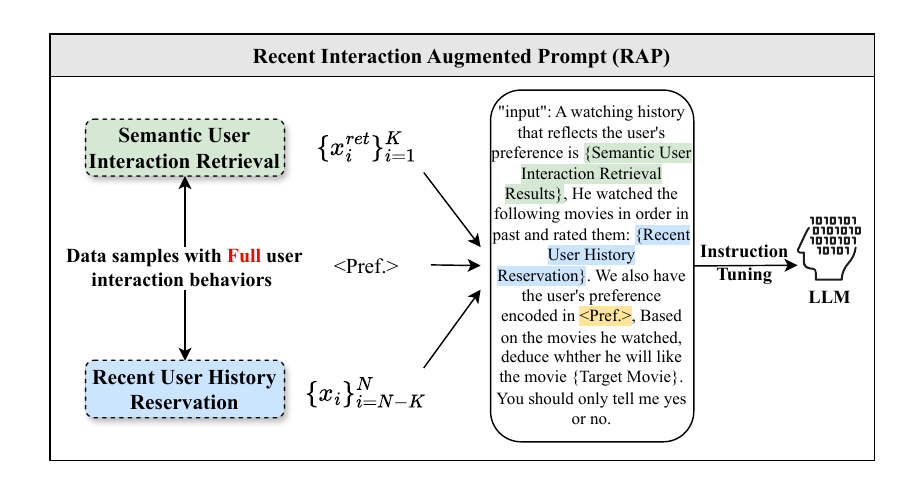}
    \vspace{-15pt}
    \caption{Construction process of RAP template. $x_i^{ret}$ denotes the retrieved history items.} 
    \label{fig:rap}
\end{figure}

Besides the retrieved sub-sequence, we further extract user's recent interacted items from the whole behavior sequence $i_1,i_2,...,i_N$, denoted as $i_{N-K}, i_{N-K+1},...i_N$, to form another sub-sequence\\$\{x_i\}_{i=N-K}^N$ of user history that contains original time series information. We design different prompts for user historical items extracted from two perspectives and integrate them into one single template. Additionally, to ensure that the RAP template can adapt to the subsequent GAT expert network's representation injection operation, we added placeholder token and design corresponding link prompts to smoothly form the whole template. An illustration of our RAP template construction process is shown in Figure \ref{fig:rap}.

\subsection{GAT Expert Network}\label{subsection: gat training}
After constructing the RAP template, this section continues to introduce the downstream GAT expert network based on it. Graph Attention Network~\citep{gat} (GAT) can efficiently perform bidirectional transmission and aggregation of node information, thereby comprehensively modeling the interaction relationships between users and items, as well as the affiliation relationships between items and features. Consequently, compared with tranditional CTR prediction models like DIN~\citep{din} and DCN~\citep{dcn}, our elaborate GAT expert network can better model the user-relecant item relationship.\par
For $i$-th user's feature $X_i^U = \langle u_{id}, u_f \rangle$, $u_{id}$ is the user ID and $u_f$ represents user features such as age and occupation. The features of the most recent $K$ items that this user has interacted with are denoted as $\{X_{i_k}^I\}_{k=N-K}^N$. Each item can be represented by a quadruple consisting of item ID, item features, user rating, and interaction timestamp, denoted as $X_{i_k}^I = \langle i_{id}^k, i_f^k, y_{i_k}, T_{i_k} \rangle$. The target item to be predicted is represented as $I_t = \langle i^t_{id}, i_f^t, T_t \rangle$. Here we use $T_*$ to represent the interaction timestamp, and for the target item, this field $T_t$ is initialized to 0.

The constructed user-item interaction graph has a set of nodes denoted as $\mathcal{V} = \{U, I_t, \{I_k\}_{k=N-K}^N, F_t, \{F_k\}_{k=N-K}^N \}$ and a set of edges represented by $\mathcal{E}=\{\{e_{U\leftrightarrow I_k}\}_{k=N-K}^N,  \{e_{I_k \leftrightarrow F_k}\}_{k=N-K}^N, e_{I_t\leftrightarrow U}, e_{I_t \leftrightarrow F_t} \}$. Namely, for the user, the target item, the $K$ most recent items and all item features, we construct a corresponding node for each of them. We also assign bidirectional edges between the user and the items, as well as between the items and their features. This forms a specific user-item-feature interaction graph $g$.\par 
The initial feature of the user node can be represented as:
\begin{equation}
    \label{equation:user rep}
    \begin{aligned}
        l_u^0 &= [Embedding(u_{id})\|Embedding(u_f)] \in \mathbb{R}^{d\times n},\\
        h_u^0 &= Linear(l_u^0) \in \mathbb{R}^{d},
    \end{aligned}
\end{equation}
where $Embedding$ stands for the embedding layer of GAT, $Linear$ is linear transform layer that maps the embedding of different kind of nodes to the same dimension $d$, $n$ is the the number of user feature fields, and $\|$ represents vector concatenation.\par
The features of an item node consist of three parts, namely ID embedding, timestamp embedding and rating embedding. The embeddings of the item ID and the rating can be directly encoded by the embedding layer, represented as:
\begin{equation}
    \label{equation:item and rating rep}
    \begin{aligned}
        l_{id}^k &= Embedding(i_{id}^k)  \in \mathbb{R}^{d}, &
        l_{y}^k &= Embedding(y_{i_k})     \in \mathbb{R}^{d},
    \end{aligned}
\end{equation}
where $i_{id}^k$ and $y_{i_k}$ stands for the item ID and numerical rating, respectively, and $l_{id}^k$ and $l_{y}^k$ is their corresponding embedding vector.

To encode interaction timestamp in Unix timestamp form, we refer to the positional encoding approach of Transformer~\citep{attention} in order to better model the time interval between current interaction time and the interaction times of each historical item. Concretely, For an item $i_k$ in the user's interaction history, its interaction timestamp is $T_{i_k}$. The gap between this timestamp and the target item's click time can be denoted as $\Delta t_k = T_t - T_{i_k}$. For the $p$-th element in the encoded embedding vector, when $p=2i$ is even, its timestamp encoding can be formulated as:
\begin{equation}
    TE_{\Delta t_k, p} = \sin (\frac{\mathrm{\Delta t_k}}{10000^{\frac{2i}{d}}}), p=2i,
\end{equation}
where $d$ is the embedding vector's dimension. And when $p=2i+1$ is odd, its timestamp encoding can be formulated as:
\begin{equation}
    TE_{\Delta t_k, p} = \cos (\frac{\mathrm{\Delta t_k}}{10000^{\frac{2i+1}{d}}}), p=2i+1.
\end{equation}

Finally, the item representation can be obtained by concatenating the embedding vectors of the above three parts and then performing linear dimensionality reduction, represented as:
\begin{equation}
   h_{i_k}^0 = Linear([l_{id}^k\|l_t^k\|l_{y}^k]) \in \mathbb{R}^d,
\end{equation}
where $h_{i_k}^0$ is the initial embedding of item node. Note that for target item, the time gap $\Delta t_k$ is set to $0$.\par

For the feature nodes in the graph, the initial representation is expressed as the result of linear dimensionality reduction of the feature embedding vector, formulated as:
\begin{equation}
    \label{equation:feature rep}
    \begin{aligned}
        l_f^k &= Embedding(f_k)  \in \mathbb{R}^{d\times n}, &
        h_{f_k}^0 &= Linear(l_f^k) \in \mathbb{R}^{d},
    \end{aligned}
\end{equation}
where $l_f^k$ denotes the embedding vector, and $h_{f_k}^0$ is the initial representation.\par

\subsection{Embedding Injection}\label{subsection: embedding injection}
Based on the GAT expert network, we further inject the embedded representations of the graph neural network into the semantic space of LLM and perform instruction tuning together with textual embedding from LLM's embedding layer.\par
A data sample from RAP template can be formulated as $p=(t,s,u,r,y)$, in which $t$ represents the share conjunction tokens in the template, $s$ and $r$ are the textual tokens of the retrieved and the recent interacted items, respectively. $y$ represents the tokens of labeled output, $u$ represents the placeholder token for the position where we are about to inject the embedding from the GAT expert network. The above text is then tokenized and encoded by the LLM's textual embedding layer, generating semantic embedding:
\begin{equation}
    \mathbf{E} = [\mathbf{e}_{t_1}, \mathbf{e}_{t_2}, \cdots, \mathbf{e}_{t_k}, \mathbf{e}_U, \cdots, \mathbf{e}_{t_{K-1}}, \mathbf{e}_{t_K}],
\end{equation}
where $\mathbf{e}_i \in \mathbb{R}^{d_{LLM} \times 1}$ is a one-dimensional vector with the same dimension as the hidden layer of the large language model, $\mathbf{e}_U \in \mathbb{R}^{d_{LLM} \times 1}$ is the vector of the placeholder token encoded by the large language model.\par
On the other hand, let the target item be $i_t$. The representation of the target item encoded by the trained GAT model is denoted as $\mathbf{e}_{i_t} \in \mathbb{R}^{d_{GAT} \times 1}$, where $d_{GAT}$ is the GAT network's embedding size. We then apply a linear mapping layer to map the target item representation to the same dimension as LLM's hidden layer. The whole process can be expressed as:
\begin{equation}
    \begin{aligned}
        \mathbf{e}_{GAT} &= g_l(\mathbf{e}_{i_t}), & \mathbf{e}_{i_t} = f_{GAT}(i_t),
    \end{aligned}
\end{equation}
where $f_{GAT}$ refers to the GAT expert network, $g_t$ represents the linear mapping layer with weight matrix $\mathbf{W}\in \mathbb{R}^{d_{GAT}\times d_{LLM}}$.\par

Finally, we feed the following embedding vector to the latter transformer layers of LLM to perform instruction tuning:
\begin{equation}
    \mathbf{E}_{hid} = [\mathbf{e}_{t_1}, \mathbf{e}_{t_2}, \cdots, \mathbf{e}_{t_k}, \mathbf{e}_{GAT}, \cdots, \mathbf{e}_{t_{K-1}}, \mathbf{e}_{t_K}].
\end{equation}

\subsection{\modelname: Training \& Inference}
This section explains the training and inference process of our proposed \modelname. The training process can be divided into two stages. We start by pretraining our GAT expert network. After constructing the feature interaction graph, we then utilize the Graph Attention Network (GAT) to train the GAT expert network. Since the constructed graph is a heterogeneous graph which contains three types of nodes: User, Item, and Feature, as well as four different types of edges: Interacted (from User to Item), ClickBy (from Item to User), BelongTo (from Item to Feature), and HasInstance (from Feature to Item). For each type of edge, the graph attention mechanism is applied to aggregate and update the nodes.\par
Concretely, given a node $i$ and its embedding at $l$-th graph attention layer, its representation is updated using graph attention mechanism, denoted by:
\begin{equation}
    \label{equation: Graph Aggregation}
        h_i^{(l+1)} = f_{GAT}(h_i^{(l)}, \{h_j^{(l)}\}), j \in \mathcal{N}_i^*, 
\end{equation}
where $f_{GAT}$ is the graph attention layer defined at Equation \ref{equ:gat defination}, $\mathcal{N}_i^*$ denotes node $i$'s neighbor under one certain edge type, $W^l$ is the weight matrix in $l$-th layer. The above formula first calculates the node similarity, and then computes the attention weight coefficients using the softmax function. The node representations are updated based on the multi-head attention mechanism, and generate next layer's representation $h_i^{l+1}$.\par
After multiple layers of graph neural network propagation, we obtain the final representations of the user and the target item, denoted as $\mathbf{e}_u$ and $\mathbf{e}_{i_t}$, respectively. The final click-through rate estimation of the model is expressed as:
\begin{equation}
    \hat{y_i}=\sigma(\frac{\mathbf{e}_u\cdot\mathbf{e}_{i_t}}{\|\mathbf{e}_u\|\times\|\mathbf{e}_{i_t}\|}),
\end{equation}
where $\sigma(\cdot)$ is the Sigmoid activation function which maps the cosine similarity to the $[0,1]$ interval. The training loss function is expressed as the cross-entropy loss between the estimated value and the true value, denoted as: 
\begin{equation}
    \mathcal{L} = -\sum_{i} (y_i \log \hat{y_i} + (1- y_i) \log(1-\hat{y_i})).
\end{equation}

Upon obtaining the pretrained GAT expert network, we train our proposed framework by first constructing the RAP prompt template using the method described in Section \ref{subsection: rap construction} for each data sample. This data sample simultaneously enters the GAT expert network to obtain user interest representation, which is then injected into the semantic space of the large language model according to the method described in Section \ref{subsection: embedding injection} for subsequent instruction fine-tuning. During the instruction fine-tuning process, we fix the main parameters of the GAT expert network and the large language model, and only update the parameters of the linear mapping layer in the injection process and the parameters of the LoRA~\citep{lora} module used for instruction tuning. We applied the causal LLM's objective for instruction tuning, which is formulated as:
\begin{equation}
\mathop{max}\limits_{\Theta}\sum_{x, y \in \mathcal{D}}\sum_{j=1}^{|y|}\log P_\Theta(y_j|x,\mathbf{E}_{hid},y_{<j}),
\end{equation}
where $\Theta$ is the trainable parameter after LoRA wrapping, $\mathcal{D}$ denotes the dataset generated by RAP construction, $\mathbf{E}_{hid}$ is the hidden representation after feature injection. We use $y_{<j}$ to represent the token before position $j$.\par
During the inference process, we fix the model parameters. For each sample, we construct the RAP prompt template and obtain the user interest representation. This representation is then injected into the large language model to obtain its output result.

\section{Experiments}
In this section, we conduct experiments to answer the following research questions (RQs).\par
\begin{itemize}
    \item[\textbf{RQ1}] How does \modelname perform against the baselines?
    \item[\textbf{RQ2}] Does the RAP template contributes to the performance?
    \item[\textbf{RQ3}] Does the GAT expert network contributes to the performance?
    \item[\textbf{RQ4}] How is the extra cost for \modelname to encode extra features?
    \item[\textbf{RQ5}] How does the training sample number affects performance?
\end{itemize}

\subsection{Setup}
\subsubsection{Datasets}
We evaluate our model on three representative public datasets. Detailed statistical feature is listed in Table \ref{tab: dataset statisitcs}.

\begin{table}[h!]
\vspace{-10pt}
\caption{Dataset statistics.}
\label{tab: dataset statisitcs}
\vspace{-5pt}
\begin{tabular}{ccccc}
\hline
Dataset       & \# Users & \# Items & \# Samples & \# Fields \\ \hline
ML-1M & $6,040$     &  $3,706$    & $970,009$     & 8 \\
ML-25M     & $162,541$    & $59,047$  & $24,187,390$   &4\\
BookCrossing       &$278,858$    & $271,376$    & $17,714$   &10 \\\hline
\end{tabular}

\vspace{-5pt}
\end{table}

\begin{itemize}[leftmargin=10pt]
\item\textbf{ML-1M}\footnote{\url{https://grouplens.org/datasets/movielens/1m/}} is a collection of 1 million movie ratings from the MovieLens website, commonly used in recommendation researches.
\item \textbf{ML-25M}\footnote{https://files.grouplens.org/datasets/movielens/ml-25m.zip} is a larger popular movie recommendation dataset containing 25 million interactions on movies from 1995 to 2019.
\item \textbf{BookCrossing}\footnote{http://www2.informatik.uni-freiburg.de/~cziegler/BX/} is a comprehensive collection of user ratings for books, including demographic information, gathered from the BookCrossing community.
\end{itemize}

For Ml-1M and ML-25M dataset, we split the training, validation and testing dataset with the ratio of 8:1:1, according to the global timestamp, and 81,920 samples are randomly selected from the ML-25M testing dataset to form our testing data. For BookCrossing dataset, we randomly split training and testing datasets based on user with the ratio 9:1, following previous works~\citep{tallrec,rella}. 

\subsubsection{Evaluation Metrics}
Three commonly used metrics are adopted to evaluate the performance: AUC (area under the ROC curve) measures model's ability to distinguish between classes, LogLoss (binary cross-entropy loss) quantifies the accuracy of predicted probabilities by penalizing false classifications, ACC (accuracy) directly calculates the proportion of correctly predicted instances.
\subsubsection{Baseline Models}
CTR prediction baseline models can be classified into three mainly categories: (1) \textbf{Feature interaction models} capture high-order relationships between user and item features by modeling collaborative signals. (2) \textbf{User behavior models} leverage user behavior sequences to capture the development trends of user interests. (3) \textbf{Semantic-enhanced models} leverages the encoding or inference capabilities of LMs to enhance recommendation performance. We choose AutoInt~\citep{autoint}, DCNv1~\citep{dcn}, DCNv2~\citep{dcnv2}, DeepFM~\citep{deepfm}, FGCNN~\citep{fgcnn}, FiGNN~\citep{fignn}, PNN~\citep{pnn}, Wide\&Deep~\citep{widedeep} and xDeepFM~\citep{xdeepfm} as the representative feature interaction models, and choose DIN~\citep{din}, DIEN~\citep{dien}, GRU4Rec~\citep{gru4rec}, Caser~\citep{caser} and SASRec~\citep{sasrec} as representative user behavior models. For semantic enhanced models, we choose P5~\citep{p5}, PTab~\citep{ptab}, CTR-BERT~\citep{CTR-BERT}, CoLLM~\citep{collm} and ReLLa~\citep{rella} as the competitive baselines.   

\subsubsection{Implementation Details}
For feature interaction and user behavior models, we train and test the models on full datasets. We add the average vector of historical item embeddings as an extra feature for the feature interaction models for fair comparison. The learning rates for these models are searched within range $\{1e-4, 3e-4, 5e-4, 1e-3\}$, and the weight decay parameters are searched within $\{1e-5, 3e-5, 5e-5, 1e-4, 3e-4\}$. For each model the embedding dim is set to $32$, the batch size is set to $1024$, and we set the patience of early stop to $10$.\par

We train our GAT expert network with two graph attention layers on the full training dataset of ML-1M and BookCrossing, and sample $1,000,000$ data points to train on ML-25M. 
Other settings are kept consistent with the traditional CTR models.

For the three LLM-based recommendation algorithms we evaluate i.e., CoLLM~\citep{collm}, ReLLa~\citep{rella}, and our proposed \modelname, we uniformly select Vicuna-13B \cite{vicuna} as the backbone LLM to ensure a fair comparison. To enhance training efficiency, we apply 8-bit quantization and use LoRA for parameter-efficient fine-tuning. The rank of the low-rank matrices in LoRA is uniformly set to $8$, alpha set to $16$, and dropout is set to $0.05$. The LoRA modules are uniformly applied to the $Q$ and $V$ parameter matrices of the attention modules in each layer of the LLM. All the three models are optimized using the AdamW~\citep{adamw} optimizer. Due to the expensive training overhead of LLM, we randomly sample $65,536$ training samples on ML-1M and ML-25M datasets, and use all $15,942$ training data on BookCrossing. 
Each model is trained for one epoch, with the learning rate fixed at $1e-3$, weight decay set to $0$, and batch size set to $256$. We only encode the item title feature to form the textual prompt template. Traditional CTR models, however, use all feature information. 

\subsection{Overall Performance (RQ1)}
In this section, we compare our proposed model \modelname with various baseline models. Their performance is listed in Table \ref{tab: main table}.
\begin{table*}
\vspace{-7pt}
\caption{Major results. The best results for each metric are given in bold, and the second best results are underlined. Rel.Impr. denotes the relative AUC improvement of \modelname w/ RAP against other baselines. The symbol $\ast$ represents significant improvement with $p$-value < 0.001. For feature interaction models we add history item embedding for a fair comparison.
}
\vspace{-5pt}
\label{tab: main table}
\resizebox{0.965\textwidth}{!}{
\renewcommand\arraystretch{1.1}
\begin{tabular}{c|c|cccc|cccc|cccc}
\toprule
\hline

\multicolumn{2}{c|}{\multirow{2}{*}{Model}} & \multicolumn{4}{c|}{ML-1M} & \multicolumn{4}{c|}{ML-25M} & \multicolumn{4}{c}{BookCrossing} \\ 
\multicolumn{2}{c|}{} & AUC  & Log Loss & ACC & Rel.Impr & AUC  & Log Loss & ACC & Rel.Impr & AUC  & Log Loss & ACC & Rel.Impr\\ 
   \hline 
   
\multicolumn{1}{c|}{\multirow{10}{*}{\makecell{Feature\\ Interaction\\ Models}}} & AutoInt &0.7952&0.5434&0.7241&2.23\%&0.8140&0.4843&0.7690&4.03\%&0.7453&0.5958&0.6710&1.11\%  \\
\multicolumn{1}{c|}{\multirow{10}{*}{}} & DCNv1 &0.7941&0.5438&0.7227&2.37\%&0.8149&0.4869&0.7692&3.91\%&0.7470&0.6011&0.6738&0.88\%  \\
\multicolumn{1}{c|}{\multirow{10}{*}{}} & DCNv2 &0.7923&0.5487&0.7196&2.60\%&0.8133&0.4854&0.7693&4.12\%&0.7469&0.5942&0.6733&0.90\%  \\
\multicolumn{1}{c|}{\multirow{10}{*}{}} & DeepFM &0.7958&0.5443&0.7223&2.15\%&0.8130&0.4859&0.7683&4.16\%&0.7458&0.5966&0.6721&1.05\%  \\
\multicolumn{1}{c|}{\multirow{10}{*}{}} & FGCNN &0.7950&0.5428&0.7230&2.25\%&0.8139&0.4860&0.7705&4.04\%&0.7400&0.6079&0.6755&1.84\%  \\
\multicolumn{1}{c|}{\multirow{10}{*}{}} & FiBiNet &0.7978&0.5476&0.7192&1.89\%&0.8151&0.4843&0.7703&3.89\%&0.6987&0.6331&0.6518&7.86\%\\
\multicolumn{1}{c|}{\multirow{10}{*}{}} & FiGNN &0.7927&0.5457&0.7207&2.55\%&0.8165&0.4841&0.7711&3.71\%&0.7408&0.6407&0.6704&1.73\%  \\
\multicolumn{1}{c|}{\multirow{10}{*}{}} & PNN &0.7943&0.5436&0.7237&2.34\%&0.8124&0.4863&0.7690&4.23\%&0.7414&0.5999&0.6761&1.65\%  \\
\multicolumn{1}{c|}{\multirow{10}{*}{}} & Wide\&Deep &0.7934&0.5453&0.7226&2.46\%&0.8126&0.4870&0.7679&4.21\%&0.7432&0.5968&0.6563&1.40\%  \\
\multicolumn{1}{c|}{\multirow{10}{*}{}} & xDeepFM &0.7967&0.5417&0.7193&2.03\%&0.8128&0.4889&0.7682&4.18\%&0.7438&0.5970&0.6704&1.32\%  \\
   \hline

\multicolumn{1}{c|}{\multirow{5}{*}{\makecell{User\\ Behavior\\Models}}} & GRU4Rec &0.7918&0.5465&0.7218&2.66\%&0.8154&0.4854&0.7680&3.85\%&0.7468&0.5981&0.6710&0.91\%  \\ 
\multicolumn{1}{c|}{\multirow{5}{*}{}} & DIN &0.7963&0.5457&0.7221&2.08\%&0.8151&0.4836&0.7701&3.89\%&0.7420&0.5966&0.6733&1.56\%  \\
\multicolumn{1}{c|}{\multirow{5}{*}{}} & DIEN &0.7970&0.5428&0.7215&1.99\%&0.8177&0.4818&0.7722&3.56\%&0.7443&0.5972&0.6653&1.25\%  \\
\multicolumn{1}{c|}{\multirow{5}{*}{}} & SASRec &0.7972&0.5403&0.7253&1.97\%&0.8187&0.4956&0.7691&3.43\%&0.7427&0.5988&0.6772&1.47\%  \\
\multicolumn{1}{c|}{\multirow{5}{*}{}} & Caser &0.7968&0.5407&0.7252&2.02\%&0.8217&0.4805&\underline{0.7745}&3.05\%&0.7462&0.5927&0.6761&0.99\%  \\

\hline
\multicolumn{1}{c|}{\multirow{6}{*}{\makecell{Semantic\\ Enhanced\\ Models}}} & P5 &0.7937&0.5478&0.7190&2.42\%&0.8126&0.5015&0.7590&4.21\%&0.7438&0.6128&0.6563&1.32\%  \\ 
\multicolumn{1}{c|}{\multirow{6}{*}{}} & PTab &0.7955&0.5428&0.7240&2.19\%&0.8170&0.4998&0.7682&3.65\%&0.7429&0.6154&0.6574&1.44\%  \\
\multicolumn{1}{c|}{\multirow{6}{*}{}} & CTR-BERT &0.7931&0.5457&0.7233&2.50\%&0.8077&0.5056&0.7611&4.84\%&0.7448&0.5938&0.6704&1.18\%  \\
\multicolumn{1}{c|}{\multirow{6}{*}{}} & CoLLM &0.7993&0.5399&0.7249&1.70\%&0.8174&0.4826&0.7679&3.60\%&\underline{0.7496}&\underline{0.5868}&0.6725&0.53\%  \\
\multicolumn{1}{c|}{\multirow{6}{*}{}} & ReLLa &\underline{0.8033}&\underline{0.5362}&\underline{0.7280}&1.20\%&\underline{0.8248}&\underline{0.4723}&0.7731&2.67\%&0.7486&0.6023&\underline{0.6775}&0.67\%  \\ 
\cline{2-14}
\multicolumn{1}{c|}{\multirow{6}{*}{}} & \modelname (w/ RAP) &\textbf{0.8129*}&\textbf{0.5240*}&\textbf{0.7375*}&-&\textbf{0.8468*}&\textbf{0.4461*}&\textbf{0.7909*}&-&\textbf{0.7536*}&\textbf{0.5859*}&\textbf{0.6787}&-  \\
   \hline  
   \bottomrule          
\end{tabular}
\vspace{-5pt}
}
\end{table*}

\begin{itemize}
[leftmargin=10pt]
    \item Our proposed RAP prompt template and \modelname framework can achieve stable performance improvements over all the baseline models on three evaluation metrics. This demonstrates that our proposed model's prompt template design for time-series completion and the method of explicitly injecting timestamp and rating information using an GAT expert network, are indeed effective and can significantly enhance the model's recommendation performance.
    \item Our proposed method further surpasses the LLM-based methos i.e., ReLLa and CoLLM. This also validates that our GAT expert network's advantage over traditional CTR models in feature encoding and information propagation via a unified graph design. This enables the model to capture the comprehensive trend of changes in user interests, thereby better modeling user interests. 
    \item In the comparison of baseline model performance, although P5, PTab, and CTR-BERT utilize LM's textual encoding capability, their performance is still sub-optimal. Meanwhile, ReLLa and CoLLM surpass other traditional recommendation models and semantic recommendation models, indicating that methods of purifying user interaction history and encoding external information can improve the LLM's ability to model user interests.
    
\end{itemize}

\subsection{Ablation Studies (RQ2 \& RQ3)}
\subsubsection{Impact of RAP Template (RQ2)} 
In this section, we verify the effectiveness of the RAP template. Concretely, we remove the RAP template within with both retrieved items and recent interacted items are integrated, only retaining retrieved items. This results in the same prompt template as ReLLa. We also apply our RAP template to ReLLa. The results are shown in Table \ref{tab:RAP}. For fairness, the two item sequences using the RAP template each takes $K=15$ items, while the sequence without the template retains $K = 30$ historical items from history retrieval. All models are trained using $65,536$ samples.

\begin{table}[h]
\centering
\vspace{-5pt}
\caption{Effect of RAP template.}
    \label{tab:RAP}
    \vspace{-5pt}
\resizebox{\linewidth}{!}{\begin{tabular}{c|ccc|ccc}
\hline
\multirow{2}{*}{Datasets} & \multicolumn{3}{c|}{ML-1M}      & \multicolumn{3}{c}{ML-25M}     \\ \cline{2-7} 
 & AUC  & Log Loss  & ACC  & AUC  & Log Loss  & ACC\\ \hline
\modelname w/ RAP&0.8129&0.5240&0.7375&0.8468&0.4461&0.7909 \\
\modelname w/o RAP&0.8108&0.5263&0.7338&0.8421&0.4512&0.7876\\
ReLLa w/ RAP&0.8083&0.5282&0.7326&0.8382&0.4568&0.7859 \\
ReLLa&0.8033&0.5362&0.7280&0.8248&0.4723&0.7731\\
\hline
\end{tabular}}
\end{table}

From the results, it can be observed that the RAP prompt template can enhance the model's understanding of user sequences solely from the data side. Moreover, our proposed \modelname itself can achieve better performance, and its performance can be further enhanced with the addition of RAP template.

\subsubsection{Impact of the GAT Expert Network (RQ3)}

In this section, we validate the effectiveness of the GAT expert network that is used to encode external numerical features. 
Specifically, we sequentially removed the item interaction timestamps and ratings from the encoding of the graph node representations and verify their impact on model performance. The corresponding results are shown in Table \ref{tab:GAT}, where TS refers to timestamp field, and R refers to rating field. We train our \modelname model without the RAP template using $32,768$ samples and $K=15$ history length on ML-1M and ML-25M datasets. For comparison purposes, we also conduct tests using the AutoInt~\citep{autoint} model without rating information on both datasets. \par

\begin{table}[h]
\centering
\caption{Ablation study on numerical fields.}
    \label{tab:GAT}
    \vspace{-5pt}
\resizebox{\linewidth}{!}{\begin{tabular}{c|ccc|ccc}
\hline
\multirow{2}{*}{Datasets} & \multicolumn{3}{c|}{ML-1M}      & \multicolumn{3}{c}{ML-25M}     \\ \cline{2-7} 
 & AUC  & Log Loss  & ACC  & AUC  & Log Loss  & ACC\\ \hline
\modelname& 0.8064&0.5334&0.7297&0.8237&0.4731&0.7765\\
\modelname (w/o TS)&0.8047&0.5348&0.7306&0.8179&0.4837&0.7698\\
\modelname (w/o TS \& R)&0.7602&0.5976&0.6855&0.8028&0.4968&0.7610\\
AutoInt (w/o TS \& R)&0.7510&0.6021&0.6841&0.6794&0.6042&0.6904\\
\hline
\end{tabular}}
\end{table}

Firstly we can observe that the encoding of the rating information has the greatest impact on model performance. This is mainly because the user's historical behavior sequence contains both positive and negative samples. Without rating information, the model cannot perceive the user's preference for history items. But even under the same experimental conditions without rating information, our model still achieves performance improvements compared to traditional recommendation models.\par
One can also observe that timestamp information contributes to model performance. The above experimental results indicate that our design of explicitly encoding timestamp and rating information into the GAT expert network indeed injects information that cannot be obtained from the textual domain into LLM, thereby improving the user interest modeling capability of \modelname.\par
\begin{figure}[h]
    \centering
    \includegraphics[width=1\linewidth,trim=0cm 0cm 0cm 0.4cm, clip]{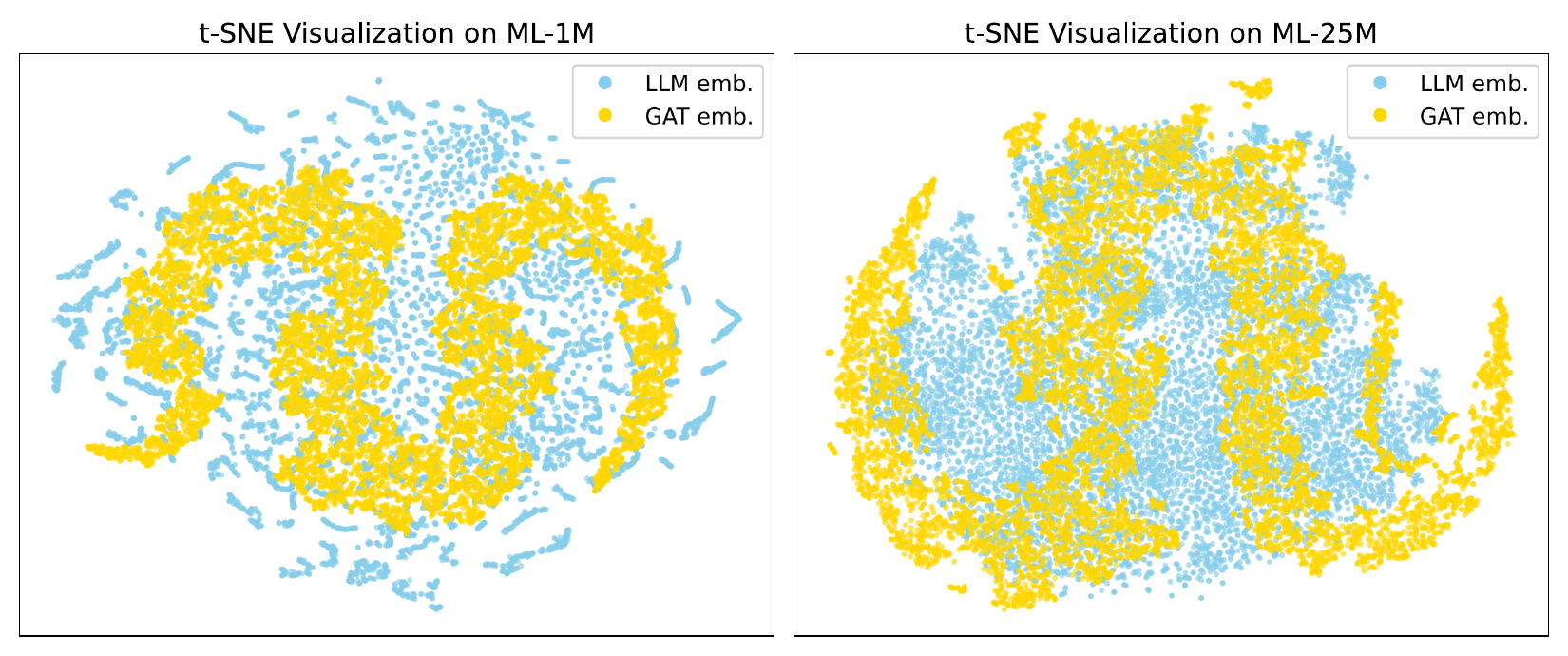}
    \vspace{-15pt}
    \caption{t-SNE visualiztion of GAT and LLM's embedding.}
    \vspace{-15pt}
    \label{fig: tsne}
\end{figure}

We also perform t-SNE~\citep{tsne} visualization on the feature vectors encoded by the GAT expert network. Specifically, we randomly select 10,000 test sample points and feed them to the trained GAT model and the text embedding layer of the large language model. We then conduct t-SNE visualization results after dimensionality reduction on the two embeddings. The visualization results on ML-1M and ML-25M datasets are shown in Figure \ref{fig: tsne}, from which we can see that the representations encoded by our GAT expert network form different manifolds in the response space, proving that they can indeed provide information that the semantic space of the large language model does not inherently possess.

\subsection{Feature Encoding Cost Analysis (RQ4)}
In this section, we validate whether our proposed model can alleviate encoding overhead. The GAT expert network has its parameters freezed during the finetune stage. \modelname incurs only a token-level cost to inject rich user historical interaction features, along with numerical rating and timestamp information that LLM finds difficult to understand, into the semantic space of it. During training, the only additional trainable parameters are the mapping layers for the GAT's embedding features. During inference, each sample only requires an extra forward pass through the GAT network. Therefore, compared to previous methods that only fine-tune a sole LLM, \modelname can more efficiently encode additional features and achieve significant improvements in recommendation performance.\par

\begin{table}[!h]
\centering
\vspace{-5pt}
\caption{Comparison of training and inference time.} 
    \label{tab:time}
    \vspace{-10pt}
\resizebox{\linewidth}{!}{\begin{tabular}{c|cc|cc}
\hline
\multirow{2}{*}{Dataset} & \multicolumn{2}{c|}{ML-1M} & \multicolumn{2}{c}{ML-25M} \\ \cline{2-5} 
                         & Training(s)    & Inference(s)    & Training(s)     & Inference(s)     \\ \hline
ReLLa                    &2.41&0.47&2.58&0.32\\
\modelname    &2.61&0.54&2.73&0.37\\ \hline
\end{tabular}}
\vspace{-5pt}
\end{table}

First we display the training and inference time of \modelname and ReLLa in Table \ref{tab:time}. The experiments are done on a single A6000 with AMD EPYC 7763 64-Core Processor. Note that for \modelname, the training and inference time includes time cost on feature encoding with the GAT expert network. From the results, we can see that though more features are encoded, our proposed method does not cause a heavy load compared with ReLLa.\par

\begin{table}[!h]
\centering
\caption{Prompt length statistic.}
    \label{tab: prompt length}
    \vspace{-5pt}
\begin{tabular}{c|cc|cc}
\hline
\multirow{2}{*}{Dataset} & \multicolumn{2}{c|}{ML-1M} & \multicolumn{2}{c}{ML-25M} \\ \cline{2-5} 
                         & Train    & Test    & Train     & Test    \\ \hline
ReLLa                    &1,484&1,517&1,515&1,557\\
\modelname    &1,676&1,548&1,730&1,748\\ 
ALL    &2,502&2,647&2,551&2,615\\ \hline
\end{tabular}
\vspace{-5pt}
\end{table}
We further count the average input prompt's length our proposed \modelname in training and testing dataset, and compare with ReLLa's input prompt and the template that directly puts all other features that our model use GAT expert network to encode into the prompt. The results are displayed in Table \ref{tab: prompt length}.\par

It can be observed from the results that our model reduces the input length by about one-quater compared to encoding all features into the input textual prompt template. 

\subsection{Data Efficiency (RQ5)}
\begin{figure}[h]
    \centering
    \includegraphics[width=1\linewidth,trim=0cm 0.4cm 0cm 0.4cm, clip]{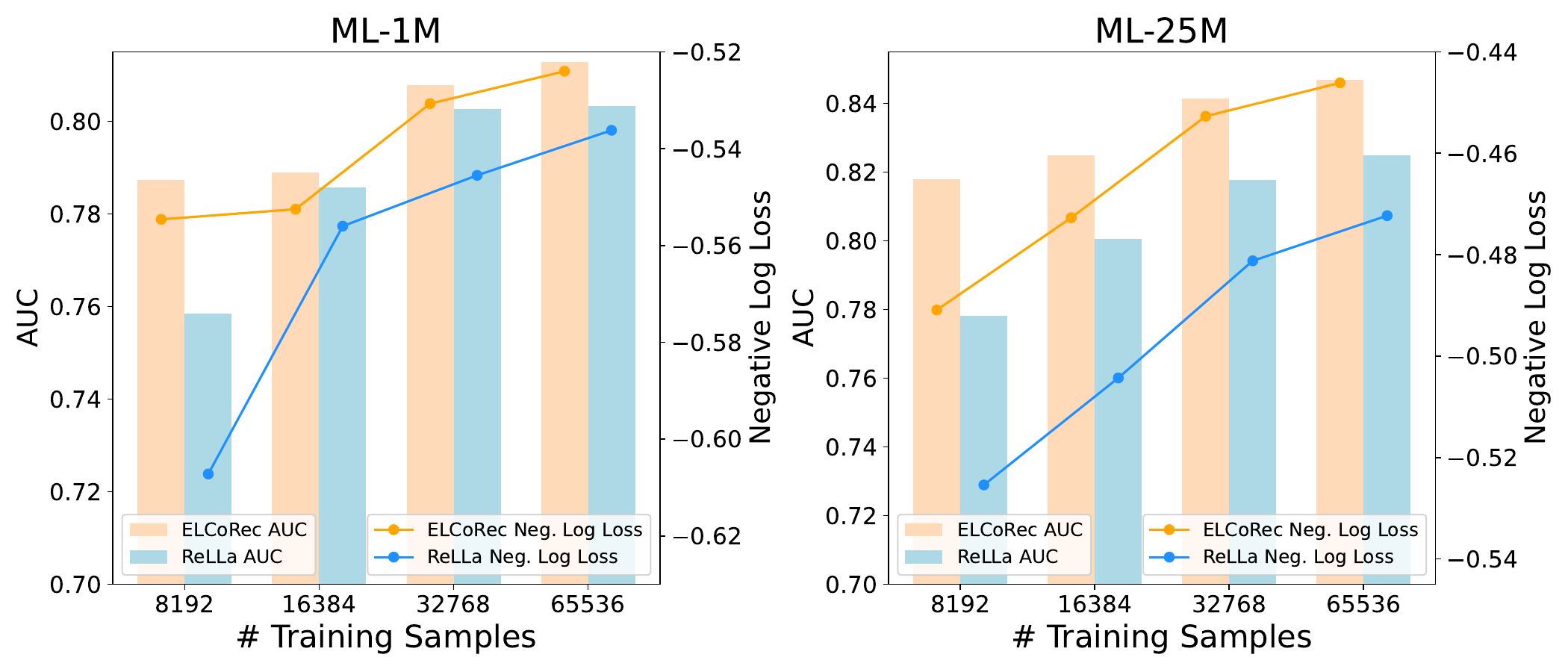}
    \vspace{-15pt}
    \caption{Data efficiency visualization.}
    \vspace{-5pt}
    \label{fig: dataeff}
\end{figure}

In this section we validate our proposed method's data efficiency by alternating the number of training samples $N$, and compare the performance with ReLLa. The results are depicted in Figure \ref{fig: dataeff}.\par


We can observe that as the number of samples $N$ increases, both ReLLa and \modelname enjoy performance improvements. However, even when training with a smaller number of samples, \modelname can still maintain higher performance than ReLLa. This indicates that our model can effectively digest and learn the different perspectives of information and knowledge brought by the prompt template and GAT expert network even under small-sample conditions, demonstrating its sample efficiency.

\section{Conclusion}
Large language models have demonstrated remarkable potential in the recommendation, but they also face challenges in comprehensively modeling user interests. In this paper, we address two critical challenges hindering the performance of LLMs in recommender systems, i.e. numerical insensitivity and encoding overhead. Our proposed framework, \modelname, along with the Recent interaction Augmented Prompt (RAP) template, provides a novel solution by incorporating recent user interactions and leveraging a graph attention network to enhance feature encoding. By injecting encoded representations into the LLM with minimal additional costs, \modelname effectively mitigates the identified issues. Extensive experiments on three public datasets validate the superior performance and efficacy of our approach. Our work sets a precedent in resolving these challenges within a unified and efficient framework, paving the way for more effective use of LLMs in recommendation tasks.



\bibliographystyle{ACM-Reference-Format}
\bibliography{sample-base}

\end{document}